\documentstyle[preprint,pra,aps]{revtex}
\tighten

\begin{document}

\def\ket#1{|#1\rangle}
\def\bra#1{\langle#1|}
\def\x{{\hat X}}
\def\y{{\hat Y}}
\def\z{{\hat Z}}
\def\H{{\hat H}}
\def\L{{\hat L}}

\preprint{NSF-ITP-98-073}

\title{Realizing the quantum baker's map on an NMR quantum computer}

\author{Todd Brun\thanks{Present address: Department of Physics, 
Carnegie Mellon University, Pittsburgh, PA 15213.
E-mail: tbrun@andrew.cmu.edu} $^{^{\hbox{\tiny(a)}}}$
and
R\"udiger Schack\thanks{E-mail: r.schack@rhbnc.ac.uk} $^{^{\hbox{\tiny (b)}}}$}

\address{$^{\hbox{\tiny (a)}}$ Institute for Theoretical Physics, \\
University of California, Santa Barbara, CA  93106, USA \\
$^{\hbox{\tiny(b)}}$ Department of Mathematics, Royal Holloway, \\ 
University of London, Egham, Surrey TW20 0EX, UK}

\date{17 December 1998}
\maketitle

\begin{abstract}
By numerically simulating an implementation of the quantum baker's map
on an NMR quantum computer based on the molecule
trichloroethylene, we demonstrate the feasibility of quantum chaos
experiments on present-day quantum computers. We give detailed
descriptions of proposed experiments that investigate (a) the rate of
entropy increase due to decoherence and (b) the phenomenon of
hypersensitivity to perturbation.
\end{abstract}

\section{Introduction}

The quantum baker's map \cite{Balazs1989,Saraceno1990} is a simple map invented
for the theoretical investigation of quantum chaos. Its mathematical properties
have been studied extensively, but no experimental quantum systems are known
which embody it.
Recently, one of us has shown that the quantum baker's map has a simple
realization on a quantum computer {\cite{Schack1998a}}. In this paper, we
present realistic numerical simulations of an NMR quantum computer
{\cite{Gershenfeld1997,Cory1997,Cory1997b}} using three quantum bits (qubits)
to explore the chaotic properties of the quantum baker's map.

In Section~\ref{secqubit} we review the definition of the quantum baker's map
and its qubit realization. We present a simple argument \cite{Mosca1997a}
showing that the quantum baker's map is equivalent to a shift map
{\cite{Alicki1994,Saraceno1994a}} on a string of qubits. This leads to the
definition of a simplified quantum baker's map which we use in the later
parts of this paper.

In Section~\ref{secnmr} we review the 3-qubit NMR quantum computer based on the
molecule trichloroethylene used in Ref.~\cite{NMRGHZ} and give the
radio-frequency (RF) pulse sequence for the proton and carbon spins of the
molecule implementing the quantum baker's map. Since decoherence cannot be
neglected for this experiment, we model the NMR system by a master equation of
Lindblad form {\cite{Lindblad1976}}, including the Hamiltonian time evolution,
the RF pulses, and phase noise due to the environment, using the actual
experimental parameters.

Finally, in Section~\ref{secchaos} we propose two specific quantum chaos
experiments. In both experiments, we compare the quantum baker's map with a
trivial {\it regular\/} map. One experiment analyses the rate of increase of
the von Neumann entropy due to decoherence \cite{Zurek1994a,Zurek1995a}, which
is related to a quantum generalization of the Kolmogorov-Sinai (KS) entropy
\cite{Alicki1994,Alicki1996b}. The other experiment examines whether the
3-qubit quantum baker's map is {\it hypersensitive to
perturbation}. Hypersensitivity to perturbation is an information-theoretic
criterion for classical and quantum chaos
{\cite{Caves1993b,Schack1996a,Schack1996b,complexity}}
which has been shown to be
equivalent to a standard definition of classical chaos under  general
assumptions {\cite{Schack1996a}}.

\section{The quantum baker's map as a shift map on qubits} \label{secqubit}

The classical baker's transformation \cite{Arnold1968}, which maps the unit
square $0 \leq q,p \leq 1$ onto itself, has a simple description in terms of
its symbolic dynamics \cite{Alekseev1981}. Each point in phase space is
represented by a symbolic string $s = \cdots s_{-2} s_{-1} s_0 . s_1 s_2
\cdots$ where $s_k=0$ or $1$. 
In the string $s$, the bits to the right of the dot are the binary expansion 
of the $q$ coordinate, and the bits to the left of the dot, read backwards,
are the binary expansion of the $p$ coordinate. Written formally, 
$s$ is identified with a point $(q,p)$
in the unit square by setting $q=\sum_{k=1}^{\infty}s_k 2^{-k}$ and
$p=\sum_{k=0}^{\infty}s_{-k} 2^{-k-1}$.  The action of the baker's map on a
symbolic string is given by the shift map $U$ defined by $(Us)_k=s_{k+1}$,
which means that, at each time step, the entire string is shifted one place to
the left while the dot remains fixed.  Geometrically, if $q$
labels the horizontal direction and $p$ labels the vertical, the
baker's map on the unit square is equivalent to stretching the $q$ direction
and squeezing the $p$ direction each by a factor of two, then
stacking the right half on top of the left.
The definition of the baker's map through its symbolic dynamics emphasizes
its prototypical character for investigations of chaotic maps: 
A very large class of chaotic maps can be shown to be 
equivalent to shifts on symbolic strings \cite{Alekseev1981}.
Furthermore, it is shown below that the quantum baker's map is equivalent to 
a shift on a string of qubits.

To define the quantum baker's map\cite{Balazs1989}, we quantize the unit square
as in \cite{Saraceno1990,Weyl1950}. To represent the unit square
in $D$-dimensional Hilbert space, we start with unitary ``displacement''
operators $\hat U$ and $\hat V$, which produce displacements in the
``momentum'' and ``position'' directions, respectively, and which obey the
commutation relation \cite{Weyl1950}
\begin{equation}
\hat U\hat V = \hat V\hat U\epsilon \;,
\end{equation}
where $\epsilon^D=1$. We choose $\epsilon=e^{2\pi i/D}$.  We further assume
that $\hat V^D=\hat U^D=1$, i.e., periodic boundary conditions. It
follows\cite{Saraceno1990,Weyl1950} that the operators $\hat U$ and $\hat V$
can be written as
\begin{equation}
\hat U=e^{2\pi i\hat q}\qquad\mbox{and}\qquad \hat V=e^{-2\pi i\hat p} \;.
\end{equation}
The ``position'' and ``momentum'' operators $\hat q$ and $\hat p$ both have 
eigenvalues $j/D$, $j=0,\ldots,D-1$.

In the following, we restrict the discussion to the case $D=2^N$, i.e., the
dimension of Hilbert space is a power of two.  For consistency of units, we let
the quantum scale on ``phase space'' be $2\pi\hbar=1/D=2^{-N}$. A
transformation between the position basis $\{|q_j\rangle\}$ and the momentum
basis $\{|p_j\rangle\}$ is effected by the discrete Fourier transform $F_N$,
defined by
\begin{equation}
F_N|q_j\rangle = |p_j\rangle = 
\sqrt{2\pi\hbar}\sum_{k=0}^{D-1}\,e^{ip_jq_k/\hbar} |q_k\rangle  =
{1\over\sqrt D}\sum_{k=0}^{D-1}\,e^{2\pi ikj/D} |q_k\rangle\;.
\label{eqfourier}
\end{equation}
The $D=2^N$ dimensional Hilbert space modeling the
unit square can be realized as the product space of $N$ qubits (i.e. $N$
two-state systems) in such a way that
\begin{equation}
|q_j\rangle =
|a_{N-1}\rangle\otimes|a_{N-2}\rangle\otimes\cdots\otimes|a_0\rangle \;,
\label{eqtensor}
\end{equation}
where $j=\sum a_k2^k$, $a_k\in\{0,1\}$ ($k=0,\ldots N-1$), 
and where each qubit has basis states $|0\rangle$ and $|1\rangle$.
It follows that, written as a binary expansion, 
$q_j=0.a_{N-1}\ldots a_0\equiv a_{N-1}2^{-1}+\cdots+a_02^{-N}$. 

There is no unique way to quantize a classical map. Here we adopt the quantized
baker's map introduced by Balazs and Voros \cite{Balazs1989}, which can be
written as \cite{Schack1998a}
\begin{equation}
T = F_N^{-1} \big(I\otimes F_{N-1}\big) \;,
\label{eqbaker}
\end{equation}
where $F_{N-1}$ acts on the $N-1$ least significant qubits, and $I$ is the
identity operator acting on the most significant qubit.  Saraceno
{\cite{Saraceno1990}} has introduced a quantum baker's map with stronger
symmetry properties by using antiperiodic boundary conditions, but in this
article we restrict the discussion to periodic boundary conditions as used
in\cite{Balazs1989}. It is straightforward to adapt the discussion in this
paper to Saraceno's version of the map {\cite{Caves1997a}}.

Note that \cite{Cleve1998} 
\begin{eqnarray}
|\psi\rangle & \equiv &
F_N^{-1}|a_{N-1}\rangle\otimes\cdots\otimes|a_0\rangle   \nonumber \\ 
& = & (|0\rangle+e^{-2\pi i(0.a_0)}|1\rangle) \otimes \cdots \otimes
%(|0\rangle+e^{-2\pi i(0.a_1a_0)}|1\rangle) \otimes \cdots \otimes
(|0\rangle+e^{-2\pi i(0.a_{N-2}\ldots a_0)}|1\rangle) \otimes
(|0\rangle+e^{-2\pi i(0.a_{N-1}\ldots a_0)}|1\rangle)
\end{eqnarray}
and
\begin{eqnarray}
|\phi\rangle & \equiv & \big(I\otimes F_{N-1}\big)^{-1}
   |a_{N-1}\rangle\otimes\cdots\otimes|a_0\rangle   \nonumber \\
& = & |a_{N-1}\rangle \otimes
 (|0\rangle+e^{-2\pi i(0.a_0)}|1\rangle) \otimes \cdots \otimes
%(|0\rangle+e^{-2\pi i(0.a_1a_0)}|1\rangle) \otimes \cdots \otimes
(|0\rangle+e^{-2\pi i(0.a_{N-2}\ldots a_0)}|1\rangle)\;.
\end{eqnarray}

As $|\psi\rangle=T|\phi\rangle$, $T$ can be seen to perform a shift of the
qubits \cite{Mosca1997a}, in which the most significant qubit of the argument
$|\phi\rangle$ is transformed in a way that depends on all the other qubits,
becoming the least significant qubit of the image $|\psi\rangle$. The quantum
baker's map (\ref{eqbaker}) is thus equivalent to a shift map on a quantum spin
chain {\cite{Alicki1994}}, in analogy to the symbolic dynamics for the
classical baker's map.

The quantum baker's map can be realized using the following basic unitary
operations or {\it quantum gates\/}: the Hadamard transform
gate $A_m$ acting on the
$m$th qubit and defined in the basis $\{|0\rangle,|1\rangle\}$ by the matrix
\begin{equation}
A_m= {1\over\sqrt2}
      \left( \begin{array}{cc}
       1 & 1 \\ 1 & -1        \end{array} \right)  \;,
\label{hadamard_gate}
\end{equation}
and the phase gate $B_{mn}(\theta)$ operating on the $m$th and $n$th qubits
and defined by
\begin{equation}
B_{mn}(\theta)\, |a_{L-1}\rangle\otimes\cdots\otimes|a_0\rangle =
  e^{i\phi_{a_ma_n}}\, |a_{L-1}\rangle\otimes\cdots\otimes|a_0\rangle \;,
\label{phase_gate}
\end{equation}
where 
\begin{equation}
\phi_{a_ma_n} = \left\{     \begin{array}{ll}
\theta      & \mbox{if $a_m=a_n=1$ ,} \\
0           & \mbox{otherwise.}
                     \end{array}
         \right. 
\end{equation}
In addition we define the gate $S_{mn}$ which swaps the qubits $m$ and $n$.

In $D=8=2^3$ dimensional Hilbert space, one iteration of the quantum baker's
map is performed by the sequence of gates
\begin{equation}
T = S_{02} A_0 B_{01}^\dagger(\pi/2) B_{02}^\dagger(\pi/4)
  A_1 B_{12}^\dagger(\pi/2) A_2 S_{01} A_0
  B_{01}(\pi/2) A_1 \;.
\label{baker_sequence}
\end{equation}

The corresponding pulse sequence on the NMR computer is quite long and
complicated (see Section~\ref{secnmr} and the Appendix). Therefore we introduce
a simplified version, $T_{\cal M}$,
of the quantum baker's map \cite{Mosca1997a}.  $T_{\cal M}$
maps each of the states
\begin{eqnarray}
&& |a_{N-1}\rangle \otimes
 (|0\rangle+e^{-2\pi i(0.a_0\ldots a_{N-2})}|1\rangle)
 \otimes (|0\rangle+e^{-2\pi i(0.a_1\ldots a_{N-2})}|1\rangle) \nonumber\\
&&    \otimes \cdots \otimes
 (|0\rangle+e^{-2\pi i(0.a_{N-2})}|1\rangle)
\end{eqnarray}
to 
\begin{eqnarray}
&& (|0\rangle+e^{-2\pi i(0.a_0\ldots a_{N-2}a_{N-1})}|1\rangle) \otimes
 (|0\rangle+e^{-2\pi i(0.a_1\ldots a_{N-2}a_{N-1})}|1\rangle) \nonumber\\
&&       \otimes \cdots \otimes
 (|0\rangle+e^{-2\pi i(0.a_{N-1})}|1\rangle) \;,
\end{eqnarray}
and is thus equivalent to $N-1$ $a_{N-1}$-controlled
rotations, a Hadamard transform on the most significant qubit, and then a cyclic
shift of the qubits.
In $D=8=2^3$ dimensional Hilbert space, one iteration of the map $T_{\cal M}$ 
is performed by the much shorter sequence of gates
\begin{equation}
T_{\cal M} = S_{01} S_{02} A_0 B_{02}^\dagger(\pi/4) B_{01}^\dagger(\pi/2) \;.
\label{mosca3bit}
\end{equation}

Like the quantum baker's map $T$, the simplified map $T_{\cal M}$ is a shift on
a string of qubits, although $T_{\cal M}$ leads to different phase relations
between the qubits. The two maps can thus be expected to have similar chaotic
behavior. We have confirmed this expectation by comparing the numerical results
of Section~\ref{secchaos} with simulations of the full quantum baker's
map $T$ \cite{unpublished}; these simulations are not included here because,
unlike the results of the present paper, they are based on unrealistic
assumptions for the experimental parameters.

\section{Implementation on an NMR quantum computer} \label{secnmr}

\subsection{The system and its Hamiltonian}

We choose for our physical system the molecule trichloroethylene
(Fig.~1), in which the nuclear spins of the hydrogen and two
carbons serve as our qubits.  These spins weakly interact with each
other on a single molecule, but are effectively shielded from the
environment by rapid tumbling.  The molecules are placed in a strong,
uniform magnetic field and subjected to RF pulses at various frequencies.

We denote by $\x$, $\y$, and $\z$ the $\sigma_x$, $\sigma_y$,
and $\sigma_z$ Pauli matrices, respectively,
and indicate with a subscript (e.g., $\x_H$) to which spin they apply.
The Hamiltonian of the three spins in the interaction picture is
\begin{equation}
\H = {j_1\over4} \z_H \z_{C_1} + {j_2\over4}(\x_{C_1}\x_{C_2}
  + \y_{C_1}\y_{C_2} + \z_{C_1}\z_{C_2}) + {j_3\over4} \z_H \z_{C_2}
  + {\delta\over2} \z_{C_2},
\label{Hamiltonian}
\end{equation}
where $j_1 \approx 203$ Hz, $j_2 \approx 102$ Hz, $j_3 \approx 10$ Hz and
$\delta \approx - 905$ Hz \cite{Knill-private}.

It is convenient to approximate
the interaction term between
the two carbons by $(j_2/4)\z_{C_1}\z_{C_2}$, omitting the $\x\x$
and $\y\y$ terms. This approximation is somewhat difficult to justify,
but greatly simplifies the description of the
quantum gates.  Generally, it works 
well if the spin precession frequency of two spins differs
by an amount large compared to the spin coupling $j$ \cite{Jones-private}.
This is true between $H$ and $C_1$ (where the approximation has already
been made), but only borderline between
$C_1$ and $C_2$.  The approximate Hamiltonian is then
\begin{equation}
\H' = {j_1\over4} \z_H \z_{C_1} + {j_2\over4}\z_{C_1}\z_{C_2}
  + {j_3\over4} \z_H \z_{C_2} + {\delta\over2} \z_{C_2},
\label{half_approx_Hamiltonian}
\end{equation}
For our numerical simulations, we will assume the form $\H'$ of the
Hamiltonian, but we have checked the dependence of our results on this
approximation.  Whenever the data curves obtained using $\H$ differ from those
obtained using $\H'$, we present both curves. 

Since $j_3$ is small compared to the other terms in the Hamiltonian, it may be
safely neglected for the design of the pulse sequences (see below). The $j_3$
term cannot be neglected, however, in the simulation of the full dynamics
including noise [see Eq.~(\ref{eqmaster})], since it is of the same order of
magnitude as the leading noise terms. Thus, for the discussion of the pulse 
sequences only, we will assume the following, further simplified form of 
the Hamiltonian,
\begin{equation}
\H'' = {j_1\over4} \z_H \z_{C_1} + {j_2\over4}\z_{C_1}\z_{C_2}
  + {\delta\over2} \z_{C_2} \;.
\label{approx_Hamiltonian}
\end{equation}

In addition to these interactions, we can apply RF pulses which rotate the
nuclear spins about the $x$ and $y$ axes.  By controlling the pulse
frequencies, these can be selectively applied to single spins (soft pulses), or
to two or three spins at once (hard pulses)\cite{Cory1997b}. 
In our simulations, we use 
instantaneous pulses, i.e., we assume that the duration of the pulses is very
short compared to the timescale of the Hamiltonian $\H''$. This assumption is
only marginally satisfied for soft pulses. The general conclusions of this
paper, however, are not affected by this approximation, since they
do not depend on the precise form of the implemented map.

For the purposes of our simulations we also assume that the RF
pulses are timed with perfect accuracy.  Unfortunately, this
is not the case in experiment, where one can expect errors
of 1--10\% or even higher \cite{Chuang-private}.
This is an additional complication,
which muddies the argument without changing its basic conclusions;
hence we neglect it.

The general form of a quantum algorithm in an NMR computer is
a sequence of pulses, causing rotations of the individual bits,
interspersed by precisely-timed delays during which the
undriven Hamiltonian couples the neighboring spins
\cite{Gershenfeld1997,Cory1997,Cory1997b}.
In describing such a sequence,
$X(\theta)$ denotes a rotation about the $x$ axis by an angle $\theta$.
This is equivalent to multiplying the state by the operator
$\exp(i\theta\x/2)$.  $Y(\theta)$ denotes a similar rotation about
the $y$ axis.  $U(t)$ indicates a delay of duration $t$, during
which the Hamiltonian $\H''$ acts.  A sequence is to be read
from right to left, i.e., the rightmost operation is performed first.
In this way, composition of the operations follows the same sense as
operator multiplication.  Subscripts indicate which spin is acted on.

The basic gates which form the algorithm are constructed
from these simple pulse sequences, as we will show below.
Note that in all cases we neglect the overall phase of the
state.  Thus, two gates will be considered equivalent if they
agree up to an overall phase.

\subsection{One-bit gates}

We are already equipped with two families of one-bit gates, the
$x$ and $y$ rotations.  Two other useful gates, however,
are lacking:  $z$ rotations and Hadamard transforms.  Fortunately,
in both cases these can be built from sequences of simple pulses
\cite{Jones1998b}.

Rotations about $z$ can be constructed from a sequence of three $x$ and
$y$ rotations \cite{Sorenson1983}:
\begin{equation}
Z(\theta) = X(-\pi/2) Y(\theta) X(\pi/2).
\end{equation}
In fact, there are several such combinations which can be used
to produce a $z$ rotation.  As gates are put together in
a quantum algorithm, it is often useful to choose their
precise form so that a certain number of pulses
combine or even cancel out,
hence simplifying the overall sequence.  Thus,
we can write any of the following:
\begin{eqnarray}
Z(\theta) = && X(-\pi/2) Y(\theta) X(\pi/2) \nonumber\\
= && X(\pi/2) Y(-\theta) X(-\pi/2) \nonumber\\
= && Y(\pi/2) X(\theta) Y(-\pi/2) \nonumber\\
= && Y(-\pi/2) X(-\theta) Y(\pi/2),
\label{zrot}
\end{eqnarray}
as convenient in constructing the algorithm.

We also need the Hadamard transform $A$ defined in (\ref{hadamard_gate}),
which can be effected by a pair of $x$ and $y$ rotations
\begin{eqnarray}
A = && Y(\pi/2) X(\pi) \nonumber\\
= && X(-\pi) Y(-\pi/2).
\end{eqnarray}
Again, the form is chosen to simplify the sequence as much as possible.

\subsection{Two-bit gates}

Given the large family of one-bit gates to choose from, we need only
a limited selection of two-bit gates.  This is good, because we have
only a limited selection to choose from.  To build our algorithm,
all we need is the phase gate $B_{ij}(\theta)$ defined in equation
(\ref{phase_gate}) which couples pairs of bits.
The phase gate between bits $i$ and $j$ can be decomposed into
\begin{equation}
B_{ij}(\theta) =
  \exp(i\theta\z_i/4) \exp(i\theta\z_j/4) \exp(i\theta\z_i\z_j/4).
\end{equation}
The $\z_i\z_j$ term is the critical one.  Transformations of that
type are produced by the Hamiltonian evolution (\ref{approx_Hamiltonian}).
However, this includes unwanted additional terms.  We
can effectively eliminate these terms by the technique of
{\it refocusing}, in which we ``undo'' the evolution of all
but the selected term in the Hamiltonian
\cite{Sorenson1983,Gershenfeld1997,Cory1997}.
Because of the anticommutation of the Pauli matrices,
\begin{equation}
\x_i \exp(i\theta\z_i\z_j ) = \exp(-i\theta\z_i\z_j) \x_i,\ \ i \ne j.
\end{equation}
Thus, if we stick an $x$ pulse in the middle of a period of Hamiltonian
evolution, it can effectively remove the unwanted terms.
\begin{eqnarray}
\x_H \exp(-i\H''\tau) \x_H \exp(-i\H''\tau) = &&
  \exp( -i \tau j_2 \z_{C_1}\z_{C_2}/2 - i\tau\delta\z_{C_2} ) \nonumber\\
\x_{C_2} \exp(-i\H''\tau) \x_{C_2} \exp(-i\H''\tau) = &&
  \exp( -i \tau j_1 \z_H\z_{C_1}/2 ).
\end{eqnarray}
We use this to build phase gates between neighboring spins.
\begin{equation}
B_{C_1H}(-\theta) = Z_H(-\theta/2) Z_{C_1}(-\theta/2) X_{C_2}(\pi)
U(\tau) X_{C_2}(\pi) U(\tau),
\end{equation}
where $\tau=\theta/2 j_1$.  Note that one can use $y$ rotations
instead of $x$, and shuffle the order of these operations; also,
one can choose any of the expressions (\ref{zrot}) for the $z$ rotations.
A very similar expression applies for $B_{C_1C_2}$,
using $X_H$ instead of $X_{C_2}$, but we
must also include an additional $Z_{C_2}$ rotation to undo the
effects of the $\delta \z_{C_2}/2$ term in the Hamiltonian.

While the phase gate is useful in itself, it can also be used
to produce the {\it controlled-not} (CNOT) gate by nesting it
between two Hadamard transforms,
$C_{ij} = A_i B_{ij}(\pi) A_i$.

\subsection{Swaps}

A disadvantage of the Hamiltonian (\ref{approx_Hamiltonian}) is that
it only couples neighboring spins.  If we wish to perform a
two-bit gate on $H$ and $C_2$, we have to swap one of them
with the central spin, $C_1$, using the swap gates $S_{C_1H}$
and $S_{C_1C_2}$.

Swaps can be built from a sequence of CNOT gates:
\begin{equation}
S_{ij} = C_{ij} C_{ji} C_{ij}.
\end{equation}
Each CNOT is composed of a phase gate and two Hadamard
transforms, as shown above, and the phase gates in turn
are built out of precisely timed Hamiltonian evolution interspersed
with RF pulses.  Thus, to swap the $H$ and $C_1$ spins requires
the sequence
\begin{eqnarray}
S_{C_1H} = &&
A_H Z_H(\pi/2) Z_{C_1}(\pi/2) X_{C_2} U(\tau) X_{C_2} U(\tau)
  A_H A_{C_1} Z_H(\pi/2) Z_{C_1}(\pi/2)
  X_{C_2} U(\tau) X_{C_2} U(\tau)\nonumber\\
&& \times A_H A_{C_1} Z_H(\pi/2) Z_{C_1}(\pi/2)
  X_{C_2} U(\tau) X_{C_2} U(\tau) A_H,
\end{eqnarray}
where $\tau = \pi/2j_1$.
The $C_1$-$C_2$ swap is similar.
Since each of the Hadamard transforms and $z$ rotations is itself
a product of several RF pulses, we see that the swap gate
is quite large and complicated.  What is more, the three phase gates
take considerable time, allowing the system to be affected by
decoherence.  The number of RF pulses can be somewhat reduced by
carefully choosing the expressions for the Hadamard and $z$ gates
to cancel as many pulses as possible, but there is nothing to be
done about the time for the phase gates.

It is clearly to our advantage to perform as few swaps as possible.
We follow this precept in designing the map $T_{\cal M}$.  As it is,
more than half the time of the algorithm is spent in swapping.

\subsection{The map $T_{\cal M}$} \label{secmosca}

In section II we defined a simplified version of the quantum
baker's map, which in the three-bit case is given by
(\ref{mosca3bit}).
This map has two important traits.  First, all of its two-bit
gates involve bit 0.  Second, the two swap gates perform a cyclic
shift of the bits, $0\rightarrow2\rightarrow1\rightarrow0$.

Since we can only couple neighboring spins, it makes the most sense
to make 0 the central bit, $C_1$.  We can then identify bit 1 with
$H$ and bit 2 with $C_2$.  In terms of the physical bits, then,
the map becomes
\begin{equation}
T_{\cal M} = S_{C_1H} S_{C_1C_2} A_{C_1} B_{C_1C_2}(-\pi/4)
  B_{C_1H}(-\pi/2).
\end{equation}
Further examination reveals that the $S_{C_1C_2}$ is
unnecessary.  The shift $1\rightarrow0$ is necessary in order to
keep bit 0 in the central position.  The labeling of bits 1 and 2,
however, is arbitrary.  We could just have easily identified bit 1
with $C_2$ and bit 2 with $H$.  Thus, we need not perform an actual
physical swap gate here; a mental re-labeling of the bits is sufficient.

This means that the labeling of the bits will switch back and forth
between even and odd iterations of the map.  Thus, the physical
realization of the map will also differ between even and odd steps.
We therefore have two alternating sequences,
\begin{eqnarray}
T_{\rm odd} = && S_{C_1H} A_{C_1}
  B_{C_1C_2}(-\pi/4) B_{C_1H}(-\pi/2) \nonumber\\
T_{\rm even} = && S_{C_1C_2} A_{C_1}
  B_{C_1H}(-\pi/4) B_{C_1C_2}(-\pi/2).
\end{eqnarray}
Each of these gates represents a given series of RF pulses
and delays.  By choosing the form of the $z$ rotations and $A$ gates
carefully, and ordering the operators for the $B$ gates appropriately,
a certain amount of cancellation is possible, simplifying the sequence
somewhat.  In this way we arrive at the following sequence of
elementary pulses:
\begin{eqnarray}
T_{\rm odd} = && X_H(-3\pi/2) Y_H(-\pi/2) Y_{C_1}(\pi/2) X_{C_1}(-\pi/2)
  Y_{C_1}(-\pi/2) U(\tau_1) X_{C_2}(\pi) U(\tau_1) X_H(-3\pi/2) \nonumber\\
&& \times X_{C_1}(-3\pi/2) Y_H(-\pi/2) Y_{C_1}(-\pi/2)
  U(\tau_1) X_{C_2}(\pi) U(\tau_1) X_H(-3\pi/2) X_{C_1}(-3\pi/2) \nonumber\\
&& \times Y_H(-\pi/2) Y_{C_1}(-\pi/2) U(\tau_1)
  X_{C_2}(\pi) U(\tau_1) X_{C_2}(\pi/2)
  Y_{C_2}(\delta\tau_1-\pi/8) \nonumber\\
&& \times X_{C_2}(\pi/2) X_H(-5\pi/4) Y_H(-\pi/2) X_{C_1}(-11\pi/8)
  Y_{C_1}(-\pi/2) U(\tau_1),
\label{eqtodd}
\end{eqnarray}
where $\tau_1 = \pi/2j_1$.  Note that because $j_1\approx 2j_2$ we can
combine the two $B$ gates into a single time delay for $T_{\rm odd}$.

This unfortunately works in exactly the wrong direction for $T_{\rm even}$,
which consequently makes the sequence somewhat longer and more complex:
\begin{eqnarray}
T_{\rm even} = && X_{C_2}(2\delta\tau_2-3\pi/2) Y_{C_2}(\pi/2)
  Y_{C_1}(\pi/2) X_{C_1}(\pi/2) Y_{C_1}(\pi/2) U(\tau_2) X_H(\pi) \nonumber\\
&& \times U(\tau_2) X_{C_1}(-3\pi/2) X_{C_2}(2\delta\tau_2-3\pi/2)
  Y_{C_1}(-\pi/2) Y_{C_2}(-\pi/2) U(\tau_2) X_{C_1}(\pi) U(\tau_2) \nonumber\\
&& \times X_{C_1}(-3\pi/2) X_{C_2}(2\delta\tau_2-3\pi/2)
  Y_{C_1}(-\pi/2) Y_{C_2}(-\pi/2) U(\tau_2) X_{C_1}(\pi) U(\tau_2) \nonumber\\
&& \times X_H(\pi/2) Y_H(-\pi/8) X_H(-\pi/2) X_{C_2}(4\delta\tau_3-5\pi/4)
  Y_{C_2}(-\pi/2) X_{C_1}(-11\pi/8) \nonumber\\
&& \times Y_{C_1}(-\pi/2) U(3\tau_3/2)
  X_H(\pi) U(5\tau_3/2),
\label{eqteven}
\end{eqnarray}
where $\tau_2 = 2\tau_1$ and $\tau_3 = \tau_1/2$.  (Recall that these
sequences should be read from right to left.)
The total delay time for $T_{\rm odd}$ is $7\tau_1$, while the delay
time for $T_{\rm even}$ is $14\tau_1$, taking twice as long.  Thus, intrinsic
noise affects the system more during even steps than odd ones.

\subsection{The regular map $T_{\cal R}$} \label{secreg}

Having defined a quantized version of a classically chaotic map,
we need to define a regular map for comparison.  This map should
take the same time per iteration as the map $T_{\cal M}$, on average,
so that it is affected equally by the intrinsic noise.

The simplest possibility would be a map that does essentially nothing.
This implies that the effects of the Hamiltonian in coupling the
spins must be suppressed.  We do this by refocusing, as described
above.
Consider the sequence
\begin{eqnarray}
T_{\cal R} =
  X_{C_1}(\pi) U(\tau_4) X_{C_1}(\pi) U(\tau_4) X_{C_1}(\pi) U(\tau_4)
  X_{C_1}(\pi) U(\tau_4) X_{C_1}(\pi) U(\tau_4) X_{C_1}(\pi) U(\tau_4)
  \times \nonumber\\
  X_{C_1}(\pi) U(\tau_4) X_{C_1}(\pi) U(\tau_4).
\end{eqnarray}
The $x$ rotations flip the sign of all terms containing $\z_{C_1}$
in the Hamiltonian, so that the effects of these terms cancel out.
That is, all but the $\delta\z_{C_2}/2$ term of (\ref{approx_Hamiltonian})
are effectively eliminated:
\begin{equation}
X_{C_1}(\pi) U(\tau) X_{C_1}(\pi) U(\tau) \approx \exp(-i\delta\tau\z_{C_2}).
\end{equation}
Thus, we see that
\begin{equation}
T_{\cal R} \approx \exp(-4i\delta\tau_4\z_{C_2}).
\end{equation}
By performing an appropriate $z$ rotation at the end we could eliminate
this cumulative rotation of the $C_2$ spin, but as it makes no difference
to the regularity of the map we leave it unaltered for simplicity.
The total time $8\tau_4$ should equal the average iteration
time of the map $T_{\cal M}$, so
$\tau_4 = 21\tau_1/16 = 21\pi/32j_1$.

\subsection{Noise and the master equation}

While the nuclear spins are fairly well isolated, this isolation
is not perfect.  Interactions with the external environment cause
the spins to deviate from strict Hamiltonian evolution, a deviation
which becomes increasingly important as the length of the NMR
calculation increases.

The effects of the environment are described by two timescales $T_1$
and $T_2$ \cite{Goldman1988}.
$T_1$ is the timescale on which the spins relax to the thermal
state.  $T_2$ is the timescale of phase decoherence,
the result of the spins becoming correlated with the
state of the environment and hence going from a superposition of states to a
statistical mixture.  For the molecule trichloroethylene, $T_1$ exceeds $T_2$
by an order of magnitude, and can be safely neglected for this problem.

We model the effect of $T_2$ by replacing the state vector $\ket\psi$
with a density matrix $\rho$ and the Schr\"odinger equation with
a Markovian master equation in Lindblad form {\cite{Lindblad1976}},
\begin{equation}
{d\rho\over dt} = - i [\H',\rho]
  + {1\over2} \sum_k 2 \left( \L_k \rho \L_k^\dagger
  - \L_k^\dagger \L_k \rho - \rho \L_k^\dagger \L_k \right) \;,
\label{eqmaster}
\end{equation}
where $\H'$ is the Hamiltonian (\ref{half_approx_Hamiltonian}) and the
$\L_k$ are a set of operators chosen to model the effects of
decoherence.  The simplest choice of $\L_k$'s which capture the
essential physics are proportional to the $\z$ operator for
each of the three spins:
\begin{eqnarray}
\L_1 = && \sqrt{\Gamma_H} \z_H, \nonumber\\
\L_2 = && \sqrt{\Gamma_{C_1}} \z_{C_1}, \nonumber\\
\L_3 = && \sqrt{\Gamma_{C_2}} \z_{C_2}.
\end{eqnarray}
The decay rate $\Gamma$ is proportional to $1/T_2$ for each spin.

Using quantum trajectory techniques \cite{Carmichael,c++},
we can solve the above master
equation numerically for various choices of $T_2$ and various
initial states.  We treat the RF pulses as instantaneous unitary
transformations, just as before.  They interrupt the continuous
master equation evolution given above.

\section{Measures of chaos} \label{secchaos}

\subsection{Rate of entropy increase}  \label{secpazur}

The Kolmogorov-Sinai (KS) entropy, equal to 1 bit per step for the classical
baker's map, measures the asymptotic rate at which information about an initial
phase-space point must be supplied in order to be keep the ability to predict
the $n$th iterate of the map to a given accuracy. Furthermore, if a stochastic
perturbation is added to the map, the KS entropy measures the average
entropy increase per step that results from averaging over the
perturbation. Both of these properties are very easily understood in the
shift-map representation of the map \cite{Alekseev1981}.

For quantum systems, a constant rate of entropy increase in the presence of
environment-induced decoherence has been proposed
{\cite{Zurek1994a,Zurek1995a}} as a signature of chaos. This rate of entropy
increase is closely related to the concept of quantum dynamical entropy
{\cite{Alicki1994,Alicki1996b}}. As in the classical case, the analysis is
greatly facilitated by the shift-map character of the quantum baker's map.  

The entropy increase in a chaotic system is due to the exponential
magnification of small-scale, local perturbations. Since perturbing all qubits
in the spin chain at the same rate corresponds to a perturbation
on all scales simultaneously, an entropy increase can be regarded as a
signature of chaos only if the qubits are perturbed at different rates. This
condition is fulfilled in our system, since the spin relaxation timescales $T_2$
for the hydrogen and carbon atoms in trichloroethylene differ by about one
order of magnitude.

In our first proposed experiment, the system is prepared in the initial pure
state
\begin{equation}
\ket{\psi_0} = \ket{0_y}\otimes\ket{0_y}\otimes\ket{0_y}
 = {1\over2\sqrt2} (\ket0+i\ket1)\otimes(\ket0+i\ket1)\otimes(\ket0+i\ket1).
\label{initial_cond}
\end{equation}
To find the system entropy,
$S(n)$, after $n$ steps ($n=1,\ldots,6$), the map $T_{\cal M}$ is iterated $n$
times by applying the pulse sequences $(\ref{eqtodd})$ and $(\ref{eqteven})$
alternately. The final density operator $\rho(n)$ is then measured using
quantum tomography {\cite{Raymer1994a,NMRGHZ}} and the entropy is determined
from $S(n)=-{\rm tr}\big[\rho(n)\log_2\rho(n)\big]$. 
Although standard NMR techniques give one the traceless part of the density
operator only, there exist methods (at least in principle)
to determine $\rho(n)$ fully, as is required 
for determining the entropy $S(n)$. One such method is to do tomography of the
quantum operation \cite{ChuangNielsen}, where the pulse sequence is applied to
different initial states.  Another possibility would be to estimate the
relative size of the traceless component by measuring the signal strength.

We have simulated this
experiment by numerically solving the master equation (\ref{eqmaster}).
The result of the simulation is shown by the data points labeled ``chaotic'' in
Fig.~2. The data points labeled ``regular'' are the results of a control
simulation using the regular map $T_{\cal R}$ defined in
Section~\ref{secreg}. In both cases, the entropy increases rapidly and
approaches the value of 3 bits, which is the maximal von Neumann entropy in
8-dimensional Hilbert space. There is no clear difference between the regular
and chaotic cases. The apparent reason is that due to the small $T_2$
timescales $1/\Gamma_{C_1}$ and $1/\Gamma_{C_2}$ two of the three qubits are
strongly perturbed.
To show that our conclusions do not depend on the assumptions leading to the
simplified Hamiltonian (\ref{half_approx_Hamiltonian}), we have repeated the
simulations leading to Fig.~2 using the full Hamiltonian (\ref{Hamiltonian}).
The data points labeled ``regular+xy'' show
the result in the regular case; in the chaotic case, the extra terms in the
Hamiltonian made no significant difference. 

If only one qubit is strongly perturbed, however, there is a clear difference
between chaotic and regular behavior. In the simulation shown in Fig.~3, both
$1/\Gamma_{C_1}$ and $1/\Gamma_{H}$ are relatively large compared to the total
delay times needed for the sequences $T_{\rm odd}$ and $T_{\rm even}$ (see Sec.\
\ref{secmosca}), i.e., only the $C_2$ spin is strongly perturbed. These $T_2$
timescales cannot be achieved with the molecule trichloroethylene, but there
may exist other molecules with the desired properties.
In this case
the ``regular+xy'' plot differs more than in Fig.~2, but is still
clearly distinguishable from the chaotic case.

In Fig.~3, the entropy increase in the chaotic case does not have a
well-defined linear regime. The reason for this is the relabeling of the qubits
at each step, which was introduced in Sec.\ \ref{secmosca} to reduce the
complexity of the pulse sequence. At alternate steps, the strongly perturbed
$C_2$ spin thus represents either qubit 1 or qubit 2. One could eliminate this
effect by performing extra physical swap operations as described in Sec.\
\ref{secmosca}. Another possibility is to introduce artificial perturbations.

One can apply an artificial perturbation to a map $T$ by adding an extra
$z$ rotation to the least significant bit, producing the perturbed map
$T' = e^{i\pi\z_2/2} T$.  At each step, one randomly chooses either
the perturbed or unperturbed map.  A density operator results from
averaging over the two possible outcomes.  This is equivalent to
applying a superoperator ${\cal P}$,
\begin{equation}
{\cal P}(\rho) = {1\over2}\big( \rho
  + e^{i\pi\z_2/2} \rho e^{-i\pi\z_2/2} \big).
\end{equation}

The results of this and the next section depend on the strength and
locality of the perturbation, but are rather insensitive to its exact form.
This is a general property of entropy measures for quantum chaos, which have 
to be defined with respect to a class of local perturbations (see, e.g.,
{\cite{Alicki1994}}). Clearly, a perturbation that commutes with the
unperturbed dynamics will not reveal any chaotic properties of the latter.

To perturb the same logical qubit at even and odd steps of the map $T_{\cal
M}$, we apply after each odd step the perturbation superoperator
\begin{equation}
{\cal P}_{\rm odd}(\rho) = {1\over2}\big( \rho + 
   e^{i\pi\z_H/2} \rho  e^{-i\pi\z_H/2} \big)
\end{equation}
and after each even step the perturbation superoperator 
\begin{equation}
{\cal P}_{\rm even}(\rho) = {1\over2}\big( \rho + 
   e^{i\pi\z_{C_2}/2} \rho  e^{-i\pi\z_{C_2}/2} \big)
\end{equation}
to the density operator $\rho$. For the regular map $T_{\cal R}$, we apply 
${\cal P}_{\rm odd}$ at each step. In an actual experiment, a convenient way
of averaging over different perturbations consists in applying selected gradient
fields \cite{Cory1998}.

The results are shown in Fig.~4. We have assumed large relaxation times
for all three spins. The extra terms in the Hamiltonian had only a slight
effect on the chaotic case, and a negligible effect on the regular case. The
simulation differentiates well between the regular and chaotic cases, and the
latter shows the expected linear increase in entropy, followed by saturation
at 3 bits. Unfortunately, this simulation assumes possibly unrealistic
relaxation times.

\subsection{Hypersensitivity to perturbation}  \label{sechyper}

Hypersensitivity to perturbation is an information-theoretic criterion for
classical and quantum chaos {\cite{Caves1993b,Schack1996a,Schack1996b}} which
has been shown to be equivalent to a standard definition of classical chaos
under general assumptions {\cite{Schack1996a}}.  Suppose a system
is perturbed, for instance by being acted on by an unknown force with
a known distribution.
Averaging over all possible perturbations causes the entropy of the
state to increase.  One can reduce this
entropy growth by obtaining information about the perturbation,
such as the actual value of the force to some precision.  By having
more information about the perturbation, the uncertainty in the state
(and hence its entropy) decreases.  To reduce the entropy
growth by an average amount $\Delta S$ requires information
$I \ge \Delta S$ about the perturbation.  In particular,
we want to know the minimum amount of information $I_{\rm min}$
needed to produce a given entropy reduction $\Delta S$.
A system is hypersensitive to perturbation if the information $I_{\rm min}$
needed to lower the system entropy increase by an
average amount $\Delta S$ is very large
compared to $\Delta S$. Precise definitions of the quantities
$I_{\rm min}$ and $\Delta S$ are given in {\cite{Schack1996b}.
For a general introduction, see \cite{complexity}.

We will show in this section that hypersensitivity to perturbation can be
detected in the 3-qubit quantum baker's map even in the presence of the actual
noise levels for trichloroethylene. As in the last section, we define perturbed
maps $T'_{\cal R}=e^{i\pi\z_H/2}T_{\cal R}$
and $T'_{\cal M}=e^{i\pi\z_H/2}T_{\rm odd}$
or $T'_{\cal M}=e^{i\pi\z_{C_2}/2}T_{\rm even}$ for odd or even steps,
respectively.  Applying randomly at each step either the perturbed or the
unperturbed map leads, after $n$ steps, to $2^n$ possible different {\it
perturbation histories}.

Due to the fast decoherence, it is necessary to limit the number of steps to 
$n=3$, corresponding to $N_{\rm hist}=2^n=8$ different perturbation histories.
The proposed experiment is to apply, each time starting from the initial state
$\ket{\psi_0}$ defined in (\ref{initial_cond}), all $N_{\rm hist}=8$
perturbation histories, to obtain the list of final density operators 
${\cal L}=\{\tilde\rho_1,\ldots,\tilde\rho_{N_{\rm hist}\}}$ by quantum 
tomography, 
and to analyze the
distribution of the $N_{\rm hist}$ density operators in density operator space.
We assume that, in a random trial, all $N_{\rm hist}$ perturbation histories 
would occur with the same probability $1/N_{\rm hist}$.
We can find the entropy
$\overline{S}_{\rm max}=-{\rm tr}\big(\tilde\rho\log_2\tilde\rho\big)$
of the {\it average} density operator
\begin{equation}
\tilde\rho={1\over N_{\rm hist}} \sum_{j=1}^{N_{\rm hist}}  \tilde\rho_j.
\end{equation}
As argued in the last section, this should grow quickly with the number
of iterations $n$.  Our simulation using the relaxation times
for trichloroethylene (Fig.~5) gave, for $n=3$, $\overline{S}_{\rm max}=2.67$ 
bits in the chaotic case, $\overline{S}_{\rm max}=2.74$ bits in the regular 
case, and $\overline{S}_{\rm max}=2.72$ bits in the regular case with $x$ 
and $y$ terms: nearly identical values. The slightly lower
entropy in the chaotic case arises because of the alternation between $T_{\rm odd}$ and $T_{\rm even}$; it is otherwise of no significance.

The values for the entropy increase alone, therefore, do not
reveal much about the distribution of the density operators in the ensemble. In
particular, they do not reveal whether
the ensemble is {\it orthogonal} (in which
case the entropy increase corresponds to purely classical information) or
nonorthogonal (corresponding to quantum information).  Obtaining 
information about which perturbation history has been realized can reduce the
entropy from $\overline{S}_{\rm max}$ to a lower value $\overline{S}$;
analyzing the dependence of $\Delta\overline{S}=\overline{S}_{\rm
max}-\overline{S}$ on the information needed gives a measure of how
nonorthogonal the ensemble is {\cite{Schack1996b}}.

We could obtain the total possible information by determining exactly
which perturbation history occurred. This corresponds to $\log_2N_{\rm hist}=3$
bits of acquired information, and would reduce us from considering the
average density operator $\tilde\rho$ to considering only a single final
density operator $\tilde\rho_j$.  However, we could also obtain {\it partial}
information about the perturbation history by partitioning the $N_{\rm hist}$ 
final density operators into $R<N_{\rm hist}$ groups, and determining 
only which group the operator was in.
Since we are actually interested in the minimum information
$I_{\rm min}$ needed to produce a given entropy reduction, we would like
to choose groupings which maximize the entropy reduction.

More precisely, consider a partitioning of the
list $\cal L$ into $R$ groups, labeled by
$r=1,\ldots,R$.  We denote by $N_r$ the number of density operators in the
$r$th group ($\sum_{r=1}^{R}N_r=N_{\rm hist}$). 
The $N_r$ density operators in the $r$th
group and their probabilities are denoted by $\rho^r_1,\ldots,\rho^r_{N_r}$ and
$q^r_1,\ldots,q^r_{N_r}$, respectively. 
In our case, all $q_i^r=1/N_{\rm hist}$. In a random trial, the system state
will be in the $r$th group with probability
\begin{equation}
p_r=\sum_{i=1}^{N_r}q^r_i = {N_r\over N_{\rm hist}} \;.
\label{eqpr}
\end{equation}
The knowledge that the system state is in group $r$ is described by the density operator
\begin{equation}
\rho_r=p_r^{-1}\sum_{i=1}^{N_r}q_i^r\rho^r_i 
= N_r^{-1}\sum_{i=1}^{N_r}\rho^r_i \;.
\label{eqrhor}
\end{equation}
We define {\cite{Schack1996b}} the system entropy conditional on being in group
$r$,
\begin{equation}
S_r = -{\rm tr} \big( \rho_r \log_2 \rho_r \big) \;,
\end{equation}
the average conditional entropy
\begin{equation}
\overline{S} = \sum_r p_r S_r = \overline{S}_{\rm max}-\Delta\overline{S} 
   \le \overline{S}_{\rm max}\;,
\end{equation}
and the average information
\begin{equation}
I = -\sum_r p_r \log_2 p_r \ge \Delta\overline{S}\;.
\label{eqdeli}
\end{equation}
The information $I_{\rm min}$ needed about the perturbation to reduce the
system entropy by an amount $\Delta S$ is now defined as the minimum of $I$
over all groupings for which $\Delta\overline{S}\ge\Delta S$, i.e., all
groupings for which the system entropy is reduced by at least $\Delta S$.

The particular case we are treating is simple enough that we could
actually try all possible groupings to find the one which minimizes $I$
for a given $\Delta S$.  In general, however, for large numbers of
iterations the number of groupings grows far too rapidly to exhaustively
consider all possibilities.  Instead, we must find an efficient 
grouping algorithm which approximates this minimum.

To find an approximation to $I_{\rm min}$ as a function of $\Delta S$, we
introduce the concept of nearly optimal groupings. Given a {\it tolerable
entropy\/} $\Delta S$, we want to partition the list of density operators $\cal
L$ into groups so as to minimize the information $I$ without violating the
condition $\Delta\overline{S}\ge\Delta S$ or $\overline{S}\le\overline{S}_{\rm
max}-\Delta S$.  To minimize $I$, it is favorable to make the groups as large
as possible. Furthermore, to reduce the contribution to $\overline{S}$ of a
group containing a given number of density operators, it is favorable to choose
density operators that are as close together as possible in some suitable sense
(see below).

To find a nearly optimal grouping into $R$ groups, we first choose $R$ density
operators at random from the list $\cal L$. Then for each of the remaining
density operators in the list, we execute the following procedure. Let $\rho_k$
be the next density operator in the list to be grouped, and let $\rho'_i$
denote the average of all density operators grouped into group $i$ so far
(i.e. excluding all those that have not yet been grouped). Then $\rho_k$ is
added to that group $j$ for which the ``distance'' 
\begin{equation}
d(\rho'_j,\rho_k) = 
- -{\rm tr}{1\over2}(\rho'_j+\rho_k)\log{1\over2}(\rho'_j+\rho_k) 
- -{1\over2}\Big[ 
- -{\rm tr}\rho_k\log\rho_k 
- -{\rm tr}\rho'_j\log\rho'_j \Big]
\end{equation}
is minimal. Of course there exist many alternative grouping algorithms, of
which we tried several, but the one described above gave consistently the best
results (i.e., the smallest $I$ for a given $\Delta S$).

Figure 5 shows $I_{\rm min}$ versus $\Delta S$
for both the chaotic and the regular case. The slope of the chaotic curve is
roughly equal to 6, i.e., about 6$n$ bits of information about the perturbation
are needed to reduce the system entropy increase by $n$ bits. In the regular
case, 1 bit of information about the perturbation is sufficient to reduce the
system entropy increase by almost 0.7 bits (0.5 bits with $\x\x$
and $\y\y$ terms). The criterion of hypersensitivity
to perturbation thus differentiates well between chaotic and regular
behavior. Furthermore, the slope of 6 in the chaotic case is not very far from
the dimension of Hilbert space, $D=8$. A slope close to $D$ is characteristic
for a random distribution of pure states in Hilbert space, and has been
conjectured to hold for chaotic quantum systems {\cite{Schack-QCM96}}. The
steep slope indicates that the ensemble is highly nonorthogonal.

\section{Conclusion}

The quantum baker's map can be implemented with present-day technology
on a 3-qubit NMR quantum computer. In order to investigate the
feasibility of quantum chaos experiments using this system, we have
numerically solved the master equation for the NMR system, including
the Hamiltonian time evolution, the RF pulses, and phase noise due to
the environment.

We have proposed and analyzed two specific quantum chaos experiments. In both
experiments, we compare the quantum baker's map with a trivial map. One
experiment analyses the increase of the von Neumann entropy due to decoherence.
We show that in principle this experiment distinguishes well between the
chaotic and regular cases, but a successful execution requires lower
decoherence rates than seem to be achievable at present.

The second proposed experiment looks for hypersensitivity to perturbation, an
information-theoretic criterion for chaos. We have shown that hypersensitivity
to perturbation can be detected in the 3-qubit quantum baker's map even in the
presence of the actual noise levels for trichloroethylene.
Using realistic estimates for the experimental parameters, our simulations show
that this criterion differentiates very well between chaotic and regular
behavior. 

We have thus shown that the quantum baker's map displays behavior of fundamental
interest even for the 8-dimensional Hilbert space of three qubits. Quantum
computers can be used to study quantum chaos under highly controlled
experimental conditions.

\acknowledgments

The authors profited from discussions with C.~M.~Caves, I.~Chuang,
J.~A.~Jones, E.~Knill, R. Laflamme, M.~Mosca, M.~Nielsen,
and W.~Zurek. Funding for RS came partially from the UK's EPSRC.
TAB was funded in part by NSF Grant No.~PHY94-07194.

\appendix

\section{}

Here we give the RF pulses corresponding to the baker's map $T$ defined
by the sequence of gates (\ref{baker_sequence}).
This unfortunately includes interactions
between non-neighboring bits.  However, one can get around this problem
by inserting an extra pair of swap gates $S_{01}$ at the beginning
and end of the iteration, making the gate sequence
\begin{equation}
T = S_{01} S_{12} A_1 B^\dagger_{01}(\pi/2) B^\dagger_{12}(\pi/4) A_0 S_{01}
  B^\dagger_{12}(\pi/2) A_2 A_1 B_{01}(\pi/2) A_0 S_{01}.
\end{equation}
The $S_{01}$ gates cancel between iterations, so one need only swap
at the beginning and end of the entire run; and since the labeling of
bits is arbitrary, these swaps can be absorbed into the process of
initial state preparation and final state tomography.  Thus, for
each iteration we perform the sequence of gates
\begin{equation}
T' = S_{12} A_1 B^\dagger_{01}(\pi/2) B^\dagger_{12}(\pi/4) A_0 S_{01}
  B^\dagger_{12}(\pi/2) A_2 A_1 B_{01}(\pi/2) A_0,
\end{equation}
which only couples neighboring bits.

Here there is no need to change representation every other step, so we
may fix a label onto the three bits.  In this case, we will identify
bit 0 with $H$, bit 1 with $C_1$, and bit 2 with $C_2$.  The sequence
of pulses corresponding to each gate
is described in detail in section III, above.  All
that is required is to combine them into a pulse sequence for the entire map:
\begin{eqnarray}
T' = && Y_{C_1}(\pi/2) X_H(\pi) X_{C_1}(\pi) U(4\tau) X_H(\pi) U(4\tau)
  Y_{C_2}(\pi/2) X_{C_2}(8\tau\delta) Y_{C_2}(8\tau\delta-\pi/2) \nonumber\\
&& \times X_{C_1}(\pi/2) X_{C_2}(\pi/2) U(4\tau) X_H(\pi) U(4\tau)
  Y_{C_1}(\pi/2) Y_{C_2}(\pi/2) X_{C_1}(\pi) X_{C_2}(\pi) U(4\tau) \nonumber\\
&& \times X_H(\pi) U(4\tau) X_{C_1}(-\pi/2) X_{C_2}(-\pi/2)
  Y_{C_1}(-3\pi/4) X_{C_1}(\pi/2) Y_{C_2}(8\delta\tau-\pi/2)
  X_{C_2}(-\pi/2) \nonumber\\
&& \times U(\tau) X_{C_2}(\pi) U(\tau) X_{C_1}(\pi/2) X_H(\pi/2)
  Y_H(\pi/4) X_H(\pi/2) Y_{C_1}(\pi/8)
  X_{C_1}(-\pi/2) U(\tau) \nonumber\\
&& \times X_H(\pi) U(\tau) X_{C_2}(\pi/2)
  Y_{C_2}(\pi/8 - 2\delta\tau) X_{C_2}(\pi/2) U(2\tau)
  X_{C_2}(\pi) U(2\tau) X_{C_1}(\pi/2) \nonumber\\
&& \times X_H(\pi/2) U(2\tau)
  X_{C_2}(\pi) U(2\tau) X_H(-3\pi/2) Y_H(-\pi/2) X_{C_1}(-3\pi/2)
  Y_{C_1}(-\pi/2) U(2\tau) \nonumber\\
&& \times X_{C_2}(\pi) U(2\tau) Y_H(\pi/2) U(2\tau) X_H(\pi)
  U(2\tau) Y_{C_2}(\pi/2) X_{C_2}(4\delta\tau-\pi/4)
  X_{C_1}(-\pi/2) \nonumber\\
&& \times Y_{C_1}(-\pi/4) X_{C_1}(-5\pi/4) Y_{C_1}(-\pi/2) U(3\tau)
  X_{C_2}(\pi) U(3\tau) Y_H(\pi/2) X_H(\pi/4),
\end{eqnarray}
where the basic timescale is $\tau = \pi/4j_1 = \tau_1/2$.

%\bibliographystyle{prsty}
%\bibliography{/home/rschack/lit/p}

%\newpage

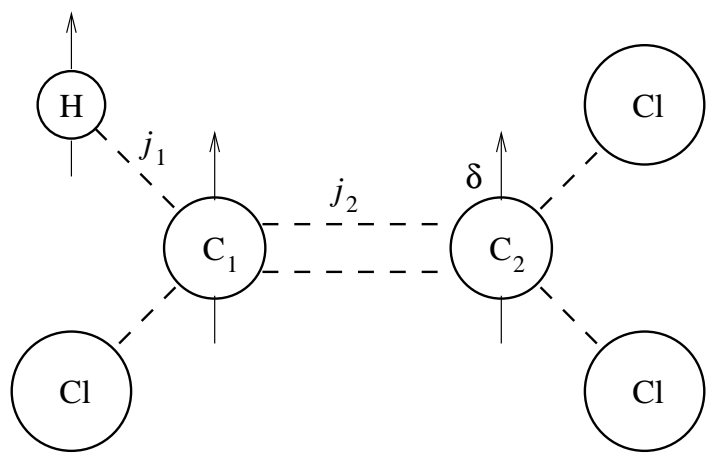
\begin{figure}
%\epsffile{fig1.ps}
\caption{The molecule trichloroethylene.}
\label{fig1}
\end{figure}

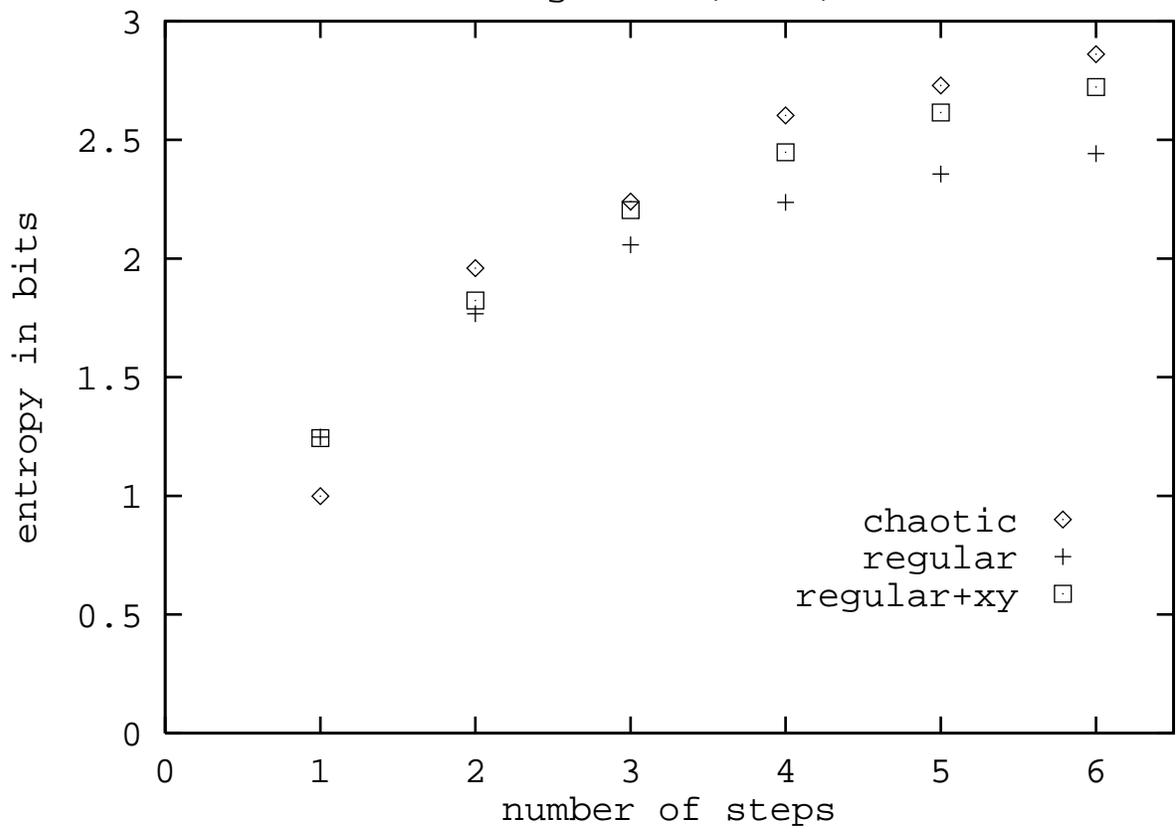
\begin{figure}
%\epsffile{fig2.ps}
\caption{Entropy versus number of steps for the regular map $T_{\cal R}$ and
the simplified baker's map $T_{\cal M}$. The decoherence times are
$1/\Gamma_{C_1}=0.7$s, $1/\Gamma_{H}=4.0$s, and $1/\Gamma_{C_2}=0.4$s, i.e.,
realistic values. The curve labeled ``regular+xy'' was generated with
the $XX$ and $YY$ terms included in the Hamiltonian. 
For the chaotic map, the effect of the extra terms is negligible.}
\label{fig2}
\end{figure}

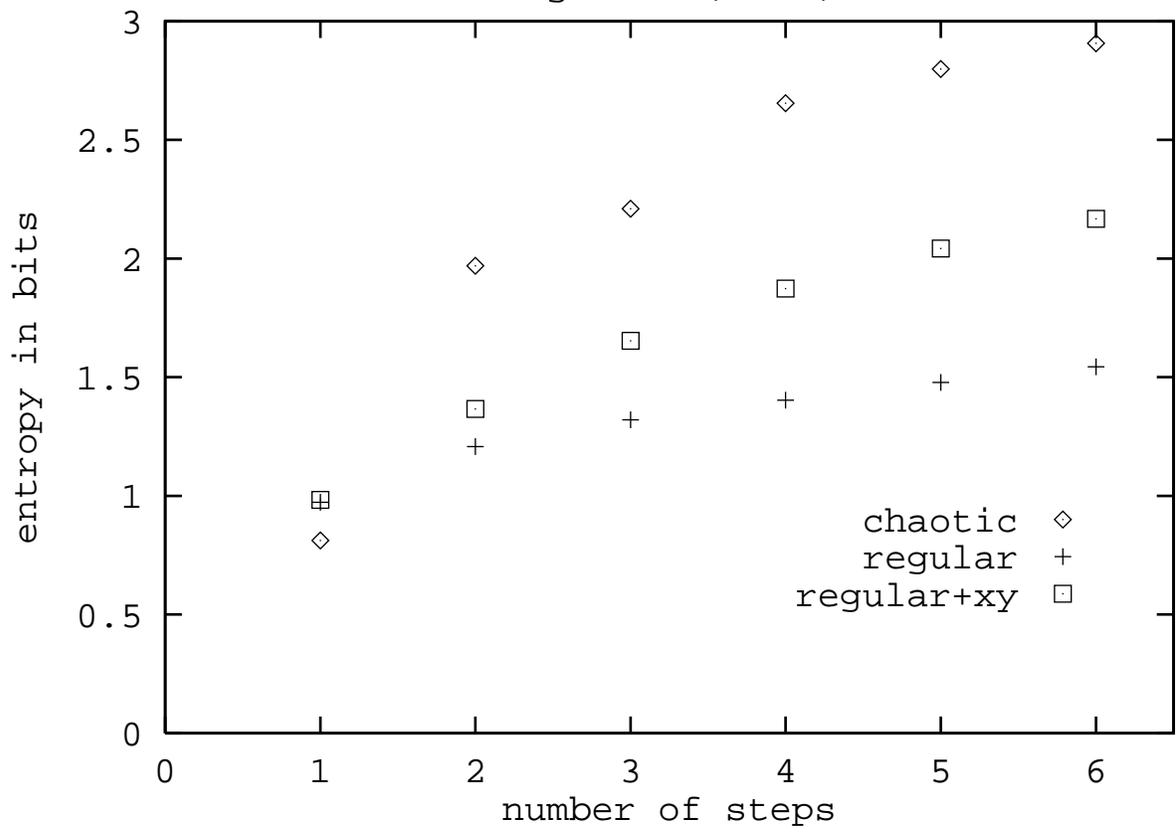
\begin{figure}
%\epsffile{fig3.ps}
\caption{Entropy versus number of steps for the regular map $T_{\cal R}$ and
the simplified baker's map $T_{\cal M}$. The decoherence times are
$1/\Gamma_{C_1}=10$s, $1/\Gamma_{H}=10$s, and $1/\Gamma_{C_2}=0.2$s, i.e.,
idealized values. The curve labeled ``regular+xy'' was generated with
the $XX$ and $YY$ terms included in the Hamiltonian. For the chaotic map, 
the effect of the extra terms is negligible.}
\label{fig3}
\end{figure}

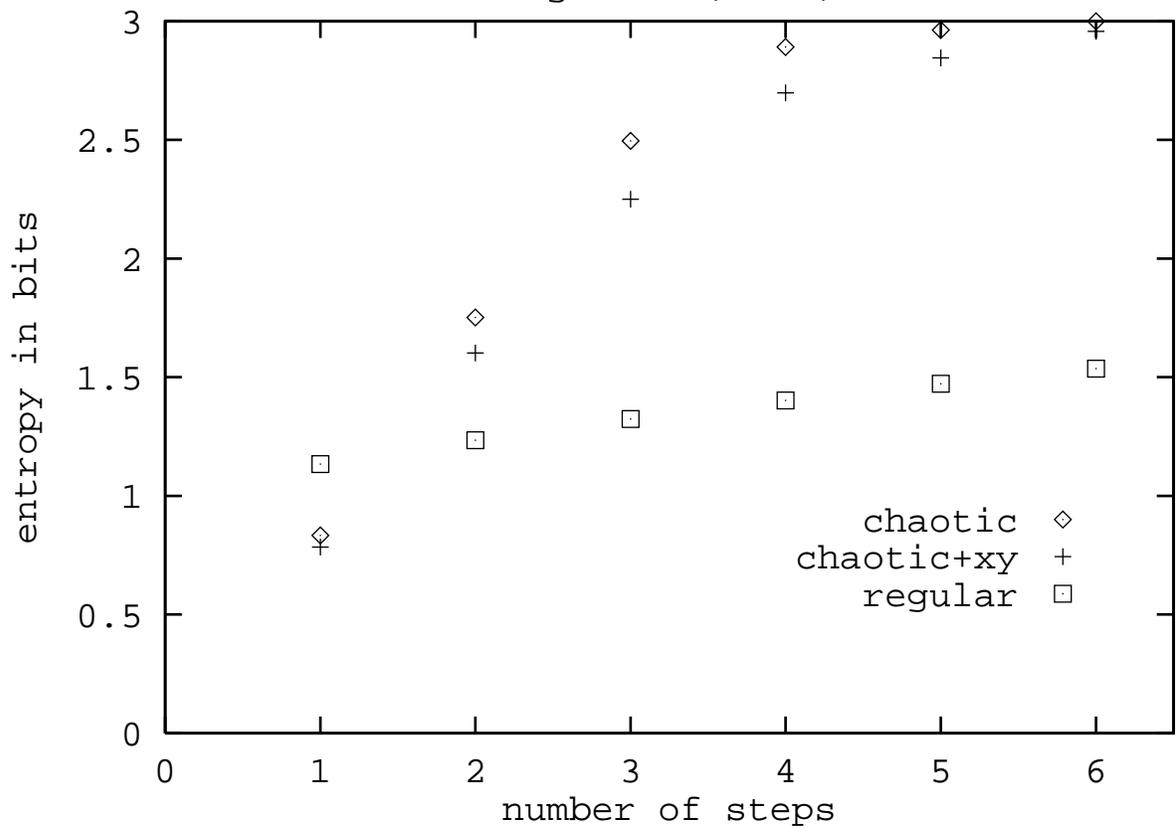
\begin{figure}
%\epsffile{fig4.ps}
\caption{Entropy versus number of steps for the regular map $T_{\cal R}$ and
the simplified baker's map $T_{\cal M}$ in the presence of an artificial
perturbation as described in the text. The decoherence times are
$1/\Gamma_{C_1}= 1/\Gamma_{H}=1/\Gamma_{C_2}=10$s, i.e., idealized values. 
The curve labeled ``chaotic+xy'' was generated with
the $XX$ and $YY$ terms included in the Hamiltonian. For
the regular map, the effect of the extra terms is negligible.}
\label{fig4}
\end{figure}

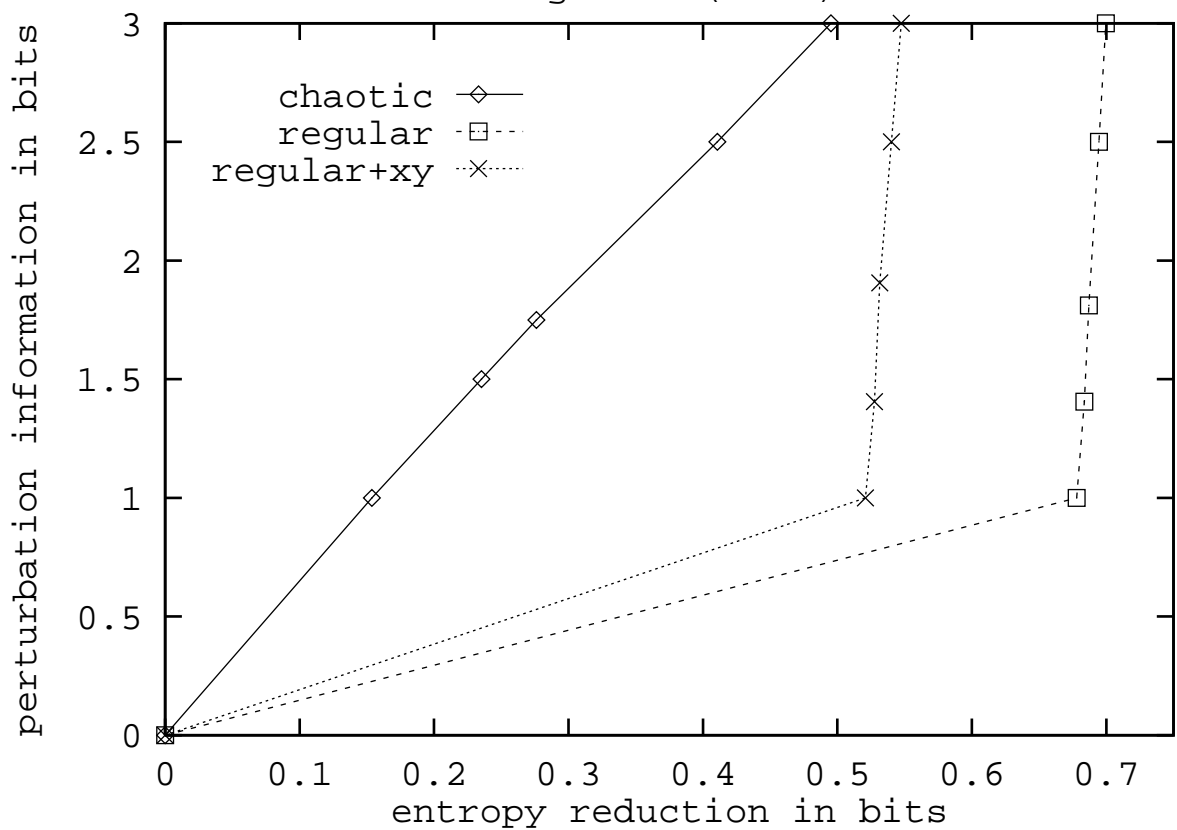
\begin{figure}
%\epsffile{fig5.ps}
\caption{Information needed about the perturbation versus entropy
reduction for the regular map $T_{\cal R}$ and
the simplified baker's map $T_{\cal M}$. The decoherence times
are $1/\Gamma_{C_1}=0.7$s, $1/\Gamma_{H}=4.0$s, and
$1/\Gamma_{C_2}=0.4$s, i.e., realistic values. 
The curve labeled ``regular+xy'' was generated with
the $XX$ and $YY$ terms included in the Hamiltonian. For the chaotic map, 
the effect of the extra terms is negligible.}
\label{fig5}
\end{figure}

\end{document}